\documentclass{llncs}
\usepackage{latexsym}
\usepackage{amssymb}
\usepackage[dvips]{graphicx}
\usepackage{float}
\usepackage{afterpage}
\usepackage{subfigure}
\usepackage{makeidx}  
\usepackage{verbatim}
\date{}

\floatstyle{ruled}
\newfloat{Program}{thp}{lop}[section]

\newfloat{algorithm}{thp}{loa}
\floatname{algorithm}{Algorithm}

\begin{document}

\title{DOEF: A Dynamic Object Evaluation Framework}

\author{\hspace{0.7cm} Zhen He$^1$  \and J{\'e}r{\^o}me Darmont$^2$}

\institute{
\begin{tabular}{cc} 
$^1$Department of Computer Science&$^2$ERIC, Universit{\'e} Lumi{\`e}re Lyon 2\\
The Australian National University&5 avenue Pierre Mend{\`e}s-France\\
Canberra, ACT 0200, Australia&69676 Bron Cedex, France\\
{\em zhen@cs.anu.edu.au}&{\em jerome.darmont@univ-lyon2.fr}\\
\end{tabular}
}

\maketitle

\vspace{-0.5cm}


\begin{abstract}

In object-oriented or object-relational databases such as multimedia databases or most XML databases, access patterns are not static,  i.e., applications do not always access the same objects in the same order repeatedly.  However, this has been the way these databases and associated optimisation techniques like clustering have been evaluated up to now.  
This paper opens up research regarding this issue by proposing a dynamic object evaluation framework (DOEF)
that accomplishes access pattern 
change by defining configurable styles of change. This preliminary prototype has been designed to be open
and fully extensible.  
To illustrate the capabilities of DOEF, we used it to compare the performances of four 
state of the art dynamic clustering algorithms.  The results show that DOEF is indeed effective at determining the 
adaptability of each dynamic clustering algorithm to changes in access pattern.  


\end{abstract}

\section{Introduction}

Performance evaluation is critical for both designers of Object Database Management Systems
(architectural or optimisation choices) and users (efficiency comparison, tuning). 
Note that we term
Object Database Management Systems (ODBMSs) both object-oriented and object-relational systems,
indifferently. ODBMSs include most multimedia and XML DBMSs, for example.
Traditionally, performance evaluation is achieved with the use of benchmarks.
While the ability to adapt to changes in access patterns is critical to database performance, none of the existing benchmarks designed for ODBMSs
incorporate the possibility of change in the pattern of object access. 
However, in real life, almost all applications do not always access the same objects in the same order repeatedly. 
Furthermore, none of the numerous studies regarding dynamic object clustering 
contain any indication of how these algorithms are likely to perform in a dynamic setting.


In contrast to the TPC benchmarks~\cite{TPC} that aim to provide \emph{standardised} means of comparing systems,
we have designed the {\em Dynamic Object Evaluation Framework} (DOEF) to provide a means to explore the performance of databases under 
\emph{different} styles of access pattern change.  

DOEF contains a set of protocols which in turn define a set of styles of access pattern 
 change.  DOEF by no means has exhausted all possible styles of access pattern change.  However, it makes 
 the first attempt at exploring the issue of evaluating ODBMSs in general and dynamic clustering algorithms
  in particular, with respect to changing query profiles.  
  
DOEF is built on top of the {\em Object Clustering Benchmark} (OCB) \cite{OCBJournal},
which is a generic benchmark that is able 
to simulate the behavior of the {\em de facto} standards in object-oriented benchmarking 
(namely OO1~\cite{OO1}, HyperModel~\cite{hypermodel}, and OO7~\cite{OO7Benchmark}).
DOEF uses both the database built from the rich schema of OCB and the operations offered by OCB.
DOEF is placed into a non-intrusive part of OCB, thus making it clean and easy to implement on 
top of an existing OCB implementation. 
Furthermore, we have designed DOEF to be open and fully extensible.  
First, DOEF's design allows new styles of change to be easily incorporated.
Second, OCB's generic model can be implemented within an object-relational system and most of its 
operations are relevant for such a system. Hence, DOEF can also be used in the object-relational context.

To illustrate the capabilities of DOEF, we benchmarked four state of the art dynamic clustering algorithms.   There are three reasons for choosing to test the effectiveness  of DOEF using dynamic clustering algorithms: ``ever since the 
early days of object database management systems, clustering has proven to be one of the most
effective performance enhancement techniques'' \cite{Gerlhof1996}; the performance of dynamic clustering algorithms 
are very sensitive to changing access patterns; and despite this sensitivity, no  previous attempt has been made to benchmark these algorithms in this way.

This paper makes two key contributions: (1) it proposes the first evaluation framework that allows ODBMS and associated optimisation techniques to be evaluated in a dynamic environment; (2) it presents the first performance evaluation experiments of dynamic clustering algorithms in a dynamic environment.

The remainder of this paper is organised as follows:
Section~\ref{sec:DOEF} describes the DOEF framework in detail,
we present and discuss experimental results achieved 
with DOEF in Section~\ref{sec:results}, and finally conclude the paper and provide future 
research directions in Section~\ref{sec:conclusion}.

\section{Specification of DOEF}
\label{sec:DOEF}


\subsection{Dynamic Framework}

We start by giving an example scenario that the framework can mimic.  Suppose we are modeling an 
on-line book store in which certain groups of books are popular at certain times.  For 
example, travel guides to Australia during the 2000 Olympics
may have been very popular.  However, once the Olympics is over, these books may suddenly or gradually
become less popular.  
Once the desired book has been selected, information relating to the book may be required.  Example 
required information includes customer reviews of the book, excerpts from the book, picture of the 
cover, etc.  In an ODBMS, this information is stored as objects referenced by the selected object 
(book), thus retrieving the related information is translated into an object graph navigation with 
the traversal root being the selected object (book).  After looking at the related information for 
the selected book, the user may choose to look at another book by the same author.  When information 
relating to the newly selected book is requested, the newly selected object (book) becomes the root 
of a new object graph traversal.
We now give an overview of the five main steps of the dynamic framework and in the process show how the above example scenario fits in.

\begin{enumerate}

\item \textbf{H-region parameters specification:} In this step we divide the database into regions of homogeneous access probability (H-regions).   In our example, each H-region represents a different group of books, each group having its own probability of access. 

\item \textbf{Workload specification:}  H-regions are responsible for assigning access probability to 
objects.  However, H-regions do not dictate what to do after an object has been selected.  We term the 
selected objects \emph{workload root}, or simply \emph{root}.
In this step, we select the type of workload to execute after selecting the root  
from those defined in OCB.  
In our example, the selected workload is an object graph traversal from the selected book to information related to the selected book, e.g., an excerpt.

\item \textbf{Regional protocol specification:} Regional protocols use H-regions to accomplish access pattern change.  Different styles of access pattern change can be accomplished by changing the H-region parameter values with time.  For example, a regional protocol may initially define one H-region with high access probability, while the remaining H-regions are assigned low access probabilities.  After a certain time interval, a different H-region may become the high access probability region.  This, when translated to the book store example, is similar to Australian travel books becoming less popular after the 2000 Olympics ends.  

\item \textbf{Dependency protocol specification:} Dependency protocols allow us to specify a relationship between the currently selected root and the next root.  In our example, this is reflected in the customer deciding to select a book which is by the same author as the previously selected book.

\item \textbf{Regional and dependency protocol integration specification:} In this step, regional and dependency protocols are integrated to model changes in dependency between successive roots.  An example is a customer using our on-line book store, who selects a book of interest, and then is confronted with a list of \emph{currently} popular books by the same author.  The customer then selects one of the listed books (modeled by dependency protocol).  The set of \emph{currently} popular books by the same author may change with time (modeled by regional protocol).
\end{enumerate}

\subsection{H-regions}
\label{sec:H-regions}

H-regions are database regions of homogeneous access probability. The parameters
that define H-regions are listed below.

\begin{itemize}
  \item \emph{HR\_SIZE:} Size of the H-region 
     (fraction of the database size).
  \item \emph{INIT\_PROB\_W:} Initial probability \emph{weight}
     assigned to the region.   The actual probability is equal to 
     this probability weight divided by the sum of all probability weights.
  \item \emph{LOWEST\_PROB\_W:} Lowest probability weight
     the region can go down to.
  \item \emph{HIGHEST\_PROB\_W:} Highest probability weight
     the region can go up to.
  \item \emph{PROB\_W\_INCR\_SIZE:} Amount by which the
    probability weight of the region increases or decreases when change is requested.   
  \item \emph{OBJECT\_ASSIGN\_METHOD:} Determines the way
     objects are assigned into the region.  
     \emph{Random} selection picks objects randomly from anywhere in the database.  By \emph{class} selection first sorts objects by class ID and then picks the first $N$ objects (in sorted order), where $N$ is the number of objects allocated to the H-region.  
  \item \emph{INIT\_DIR:} Initial direction that the probability
      weight increment moves in.  
\end{itemize}

\subsection{Regional Protocols}

Regional protocols simulate access pattern change by first initialising the parameters of every H-region, 
and then periodically changing the parameter values in certain predefined ways.  This paper documents three 
styles of regional change.   For every regional protocol, a user defined parameter \emph{H} is used to control the rate at which access pattern changes.  More precisely, \emph{H} is defined as one divided by the number of transactions executed between each change of access pattern.
Three regional protocols are listed below:
\begin{itemize}
\item \textbf{Moving Window of Change Protocol:}
This regional protocol simulates sudden changes in access pattern.  In our on-line book store, this is 
translated to books suddenly becoming popular due to some event (e.g., a TV show). Once the event passes, the books 
become unpopular very fast.  
This style of change is accomplished by moving a window through the database.  The objects in the window 
have a much higher probability of being chosen as root when compared to the remainder of the database.  
This is done by breaking up the database into \emph{N} H-regions of equal size.  Then, one H-region is 
first initialised to be the hot region (i.e., a region with high probability of reference), and after a certain number of root selections, a different H-region becomes the hot region. 

\begin{comment}
\begin{itemize}

\item The database is broken up into \emph{N} regions of equal size.
\item Set the \emph{INIT\_PROB\_W} of one of the H-regions to equal
\emph{HIGHEST\_PROB\_W} (the hot region) and the rest of the H-regions
get their \emph{INIT\_PROB\_W} assigned to \emph{LOWEST\_PROB\_W}.  
\item Set \emph{PROB\_INCR\_SIZE} value of all every region to equal \emph{HIGHEST\_PROB\_W} - \emph{LOWEST\_PROB\_W} and all H-regions are assigned the same \emph{LOWEST\_PROB\_W} and \emph{PROB\_W\_INCR\_SIZE} values.
\item Introduce a user defined parameter \emph{H} which controls the rate of access pattern change.
\item The \emph{INIT\_DIR} parameter of all the H-regions are set to move downwards.  Initially, the window is placed at the hot region.  After every 1 / \emph{H} root selections, the window moves from one H-region to another.  The H-region that the window is moving \emph{from} has its direction set to \emph{down}.  The H-region that the window is moving \emph{into} has its direction set to \emph{up}.  Then, probability weights of the H-regions are incremented or decremented depending on the current direction of movement.
\end{itemize}
\end{comment}

\item \textbf{Gradual Moving Window of Change Protocol:}
This protocol is similar to the previous one, but the hot region cools down gradually instead of suddenly.  The cold regions also heat up gradually as the window is moved onto them.  This tests the dynamic clustering algorithm's ability to adapt to a more moderate style of change.  In our book store example, this style of change may depict travel guides to Australia gradually becoming less popular after the Sydney 2000 Olympics.  As a consequence, travel guides to other countries may gradually become more popular.  Gradual changes of heat may be more common in the real world.  This protocol is implemented in the same way as the previous protocol except the H-region that the window (called the hot region in the previous protocol) moves \emph{into} gradually heats up and the H-region that the window moves \emph{from} gradually cools down.

\begin{comment}
This protocol is specified in the same way as the previous protocol with two exceptions.  Firstly, \emph{PROB\_W\_INCR\_SIZE} is now user-specified instead of equaling \emph{HIGHEST\_PROB\_W} - \emph{LOWEST\_PROB\_W}.  The value of \emph{PROB\_W\_INCR\_SIZE} determines how vigorously access pattern changes at every change iteration.  The second exception is in the way the H-regions change direction.   The H-region that the window moves \emph{into} has its direction \emph{toggled}.  The direction of the H-region that the window is moving \emph{from} is \emph{unchanged}.  This way, the previous H-region is able to continue cooling down gradually or heating up gradually. 
\end{comment}

\item \textbf{Cycles of Change Protocol:} This style of change mimics something like a bank where customers
  in the morning tend to be of one type and in the afternoon of another type.  This, when repeated, creates a cycle of change.   This is done by break up the database into three H-regions.   The first two H-regions represent objects going through the cycle of change.  The third H-region represent the remaining unchanged part of the database.  The first two H-regions alternates at being the hot region.

\begin{comment}
\begin{itemize}
  \item Break up the database into three H-regions.  The first two H-regions represent objects going through the cycle of change.  The third H-region represent the remaining unchanged part of the database.  The \emph{HR\_SIZE} of the first two H-regions are equal to each other and user-specified.  The \emph{HR\_SIZE} of the third H-region is equal to the remaining fraction of the database.
\item Set the \emph{LOWEST\_PROB\_W} and \emph{HIGHEST\_PROB\_W} parameters of the first two H-regions to values that reflect the two extremes of the cycle. 
\item Set the \emph{PROB\_INCR\_SIZE} of the first two H-regions to equal \emph{HIGHEST\_PROB\_W} - \emph{LOWEST\_PROB\_W}.  Set the \emph{PROB\_INCR\_SIZE} of the third H-region to equal zero.
 \item The \emph{INIT\_PROB\_W} of the first H-region is set to \emph{HIGHEST\_PROB\_W} and the second to \emph{LOWEST\_PROB\_W}.
  \item Set the \emph{INIT\_DIR} of the hot H-region to down and the \emph{INIT\_DIR} of the cold H-region to up. 
 \item Again, the \emph{H} parameter is used to vary the rate of access pattern change.
\end{itemize}
\end{comment}

\end{itemize}

\subsection{Dependency Protocols}

There are many scenarios in which a person executes a query and then decides to execute another query based on the results of the first query, thus establishing a dependency between the two queries.  In this paper, we have specified four dependency protocols.  All four protocols functions by finding a set of candidate objects that maybe used as the next root.  Then a random function is used to select one object out of the candidate set.  The selected object is the next root.  An example random function is a skewed random function that selects a certain subset of candidate objects with a higher probability than others.  
The four dependency protocols are listed below:
\begin{itemize}
\item \textbf{Random Selection Protocol:} This method simply uses some random function to select the current root.  This protocol mimics a person starting a completely new query after finishing the previous one.



\item \textbf{By Reference Selection Protocol:} The current root is chosen to be an object referenced by the previous root.  An example of this protocol in our on-line book store scenario is a person having finished with a selected book, who then decides to look at the next book in the series.



\item \textbf{Traversed Objects Selection Protocol:} The current root is selected from the set of objects that were referenced in the previous traversal.  An 
example is a customer requesting in a first query a list of books along with their author and publisher, 
who then decides to read an exerpt from one of the books listed.



\item \textbf{Same Class Selection Protocol:}
The currently selected root must belong to the same class as the previous root.  Root selection is further
restricted to a subset of objects of the class.  The subset is chosen by a function that takes the previous root as a parameter.  That is, the subset chosen 
dependent on the previous root object.     An example of this protocol is a customer deciding to select a book from our on-line book store which is by the same author as the previous selected book.  In this case, the same class selection function returns books by the same author.
\begin{comment}

$r_{i+1}$ = $RAND\emph{4}(f(r_{i}, Class(r_{i}), U))$

Where $Class(r_i)$ returns the class of the $i^{th}$ root. The parameter $U$ is user-defined and specifies the
size of the set returned by function $f()$.  $U$ is specified as a
fraction of the total class size.  $U$ can be used to increase or decrease the degree of locality between the objects returned by $f()$.  $f()$ always returns the same set of objects given the same set of parameters.
\end{comment}

\end{itemize}

\subsubsection{Hybrid Setting.}
\label{sec:hybrid}
The hybrid setting allows an experiment to use a mixture of the dependency protocols outlined above.  Its use is important since it simulates a user starting a fresh random query after having followed a few dependencies.  Thus, the hybrid setting is implemented in two phases.  The first \emph{randomisation phase} uses the random selection protocol to randomly select a root.  In the second \emph{dependency phase}, one of the dependency protocols outlined in the previous section is used to select the next root.  $R$ iterations of the second phase are repeated before going back to the first phase.  The two phases are repeated continuously. 


\subsection{Integration of Regional and Dependency Protocols}
\label{sec:reg_dep_prot}

Dependency protocols model user behavior.   Since user behavior can change with time, dependency protocols should also be able to change with time.  The integration of regional and dependency protocols allows us to simulate changes in the dependency between successive root selections.  This is easily accomplished by exploiting the dependency protocols' property of returning a candidate set of objects when given a particular previous root.  Up to now, the next root is selected from the candidate set by the use of a random function.  Instead of using the random function, we partition the candidate set using H-regions and then apply regional protocols on these H-regions.  

\section{Experimental Results}
\label{sec:results}

\subsection{Experimental Setup}
\label{sec:dynamic_clustering_results}
We used DOEF to compare the performance of four state of the art dynamic clustering 
algorithms: Dynamic, Statistical, and Tunable Clustering (DSTC) \cite{ecoop96_bullat}, 
Detection \& Reclustering of Objects (DRO) \cite{DRO}, dynamic Probability Ranking Principle (PRP) \cite{OPCF}, 
and dynamic Graph Partitioning (GP) \cite{OPCF}.  
The aim of dynamic clustering is to automatically place objects that are likely to be accessed 
together in the near future in the same disk page, thereby reducing the number of I/O. 
\begin{comment}
DSTC achieves dynamicity without adding high
statistics collection overhead and excessive volume of statistics. However, its clustering overhead is high.
DRO is partly based on DSTC, but 
produces less clustering I/O overhead and uses less statistics.
Finally, the opportunistic prioritised clustering framework (OPCF) transforms static clustering algorithms into dynamic clustering algorithms.
We have used it to transform the {\it static greedy graph partitioning} and {\it static probability ranking principle} clustering algorithms into dynamic clustering algorithms.
\end{comment}
The four clustering techniques have been parameterized for the same behaviour and best performance. 

We chose simulation for these experiments, principally because
it allows rapid development and testing of a large number of dynamic clustering algorithms (all previous dynamic clustering papers compared at most two algorithms).  
The experiments were conducted on the Virtual Object-Oriented Database discrete-event simulator (VOODB) \cite{voodb}.  
Its purpose is to allow performance evaluations of OODBs in general, and optimisation methods like clustering in particular.   
VOODB has been validated for real-world OODBs  in a variety of situations.
The VOODB parameter values we used are depicted in Table~\ref{table:ocb_voodb_param} (a).

Since DOEF uses the OCB database and operations, it is important for us to document the OCB settings 
used for these experiments (Table~\ref{table:ocb_voodb_param} (b)).  The size of the objects used varied 
from 50 to 1600 bytes, with an average of 233 bytes.  A total of 100,000 objects were generated for a total database size of 23.3 MB.   
Although this is a small database size, we also used a small buffer size (4 MB) to keep the database to buffer size ratio large.  Clustering algorithm performance is indeed more sensitive to database to buffer size ratio than database size alone.
 The operation used for all the experiments was a simple, depth-first traversal of depth 2.  
 We chose this simple traversal because it is the only one that always accesses the same set of objects given a particular root.  This establishes a direct relationship between varying root selection and changes in access pattern.  Each experiment involved executing 10,000 transactions.
The main DOEF parameter settings used in this study are shown in Table~\ref{table:doef_param}.  
These DOEF settings are common to all experiments in this paper.  The \emph{HR\_SIZE} setting of 0.003 creates a hot region about 3\% the size of the database.  This fact was verified from statistical analysis of the trace generated.  The $HIGHEST\_PROB\_W$ setting of 0.8 and $LOWEST\_PROB\_W$ setting of 0.0006 produce a hot region with 80\% probability of reference, the remaining cold regions having a combined reference probability of 20\%.  These settings are chosen to represent typical database application behaviour~\cite{gray87,carey91,franklin93}.


\vspace{-0.4cm}

\begin{table*}[hbt]
  \begin{center}    
  \subfigure[VOODB parameters]{
    \begin{tabular}{|l|l|}\hline
      {\bf Parameter Description} & {\bf Value}\\\hline\hline
      System class & Centralized \\\hline
      Disk page size & 4096 bytes\\\hline
      Buffer size & 4 MB  \\\hline
      Buffer replacement & LRU-1\\
      policy &\\\hline
      Pre-fetching policy & None \\\hline
      Multiprogramming level & 1\\\hline
      Number of users & 1\\\hline
            Object initial placement & Sequential\\\hline
    \end{tabular}}\quad
   \subfigure[OCB parameters]{
      \begin{tabular}{|l|l|} \hline
        {\bf Parameter Description} & {\bf Value} \\\hline\hline
        Number of classes& 50\\\hline
        Maximum number of & 10\\ 
        references, per class &\\\hline
        Instances base size, per class & 50\\\hline
        Total number of objects & 100000\\\hline
        Number of reference types & 4 \\\hline
        Reference types distribution & Uniform\\\hline
        Class reference distribution & Uniform\\\hline
        Objects in classes distribution & Uniform\\\hline
        Objects references distribution & Uniform\\\hline
    \end{tabular}} 
  \caption{VOODB and OCB parameters}
  \label{table:ocb_voodb_param}
\end{center}    
\end{table*}

\vspace{-1.5cm}

\begin{table*}[hbt]
  \begin{center}  
	\begin{tabular}{|l|l|} \hline
        {\bf Parameter Name} & {\bf Value} \\\hline\hline
	  \emph{HR\_SIZE} & 0.003\\\hline
	  \emph{HIGHEST\_PROB\_W} & 0.80\\\hline
	  \emph{LOWEST\_PROB\_W} & 0.0006\\\hline	
	  \emph{PROB\_W\_INCR\_SIZE} & 0.02\\\hline
	  \emph{OBJECT\_ASSIGN\_METHOD} & Random object assignment\\\hline
        \end{tabular}
   \caption{DOEF parameters}
   \label{table:doef_param}
  \end{center}
\end{table*}

\vspace{-1.5cm}

\begin{comment}
The dynamic clustering algorithms shown on the graphs in this section are labeled as follows:
    \textbf{NC}: No Clustering;
    \textbf{DSTC}: Dynamic Statistical Tunable Clustering;
    \textbf{GP}: OPCF (greedy graph partitioning);
    \textbf{PRP}: OPCF (probability ranking principle);
    \textbf{DRO}: Detection \& Reclustering of Objects.
    
\begin{comment}
\begin{center}
  \parbox{13cm}{
    \vspace{-2.0ex}
    \begin{itemize}
      \setlength{\itemsep}{0ex}
    \item\textbf{NC}: No Clustering;
    \item\textbf{DSTC}: Dynamic Statistical Tunable Clustering;
    \item\textbf{GP}: OPCF (greedy graph partitioning);
    \item\textbf{PRP}: OPCF (probability ranking principle);
    \item\textbf{DRO}: Detection \& Reclustering of Objects.
    \end{itemize}
    }
    \vspace{-2ex}
\end{center}
\end{comment}

As we discuss the results of these experiments, we focus our discussion on the relative ability of each algorithm to adapt to changes in access pattern, i.e., as rate of access pattern change increases, we seek to know which algorithm exhibits more rapid performance deterioration.  This contrasts from discussing which algorithm gives the best absolute performance.  All the results presented here are in terms of total I/O (transaction I/O plus clustering I/O).




\subsection{Moving and Gradual Moving Regional Experiments}

\begin{figure*}[hbt]
  \begin{center}
  \subfigure[Moving Window of Change]{\includegraphics[width = 5.85 cm]{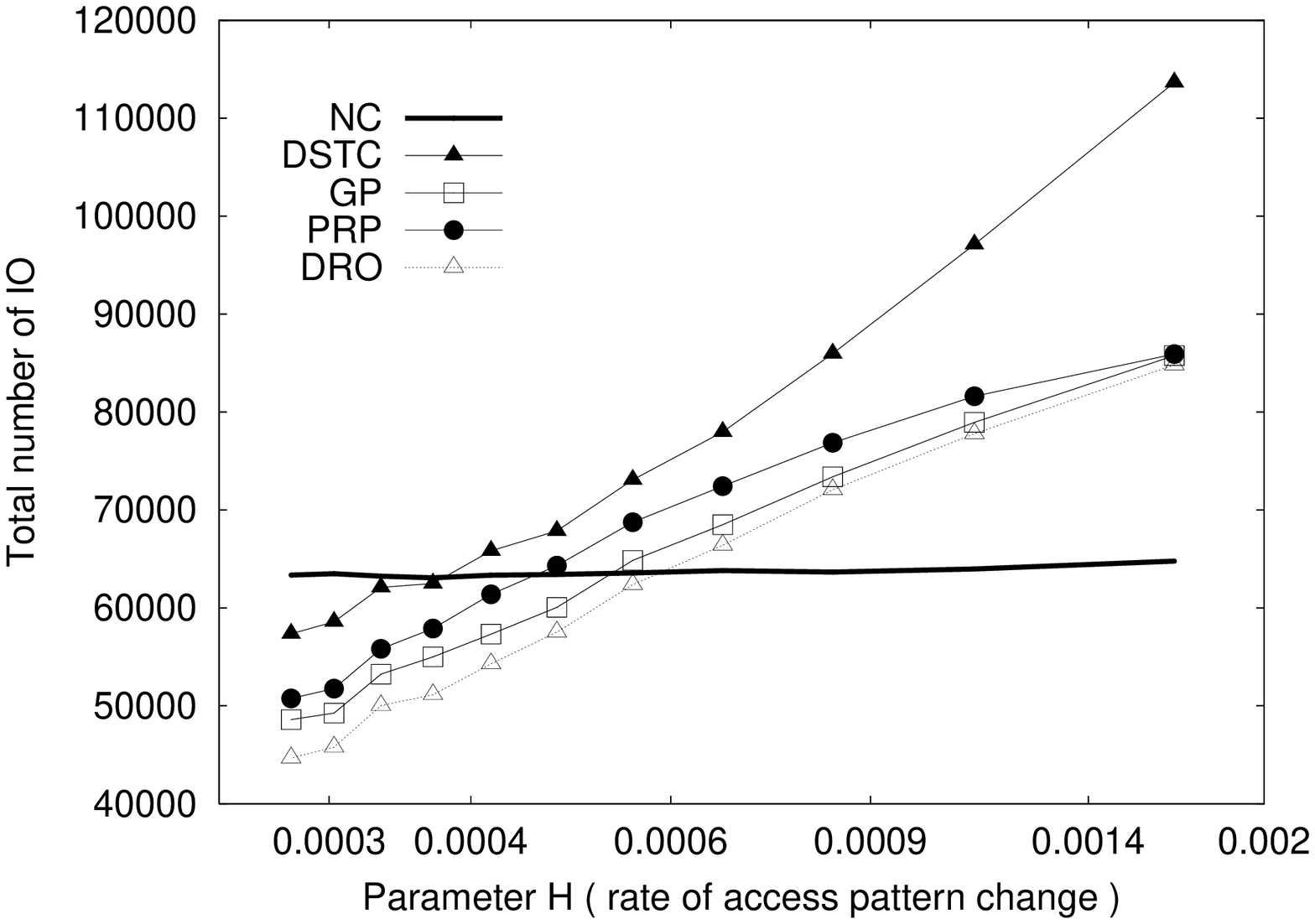}}\quad
  \subfigure[Gradual Moving Window of Change]{\includegraphics[width = 5.85 cm]{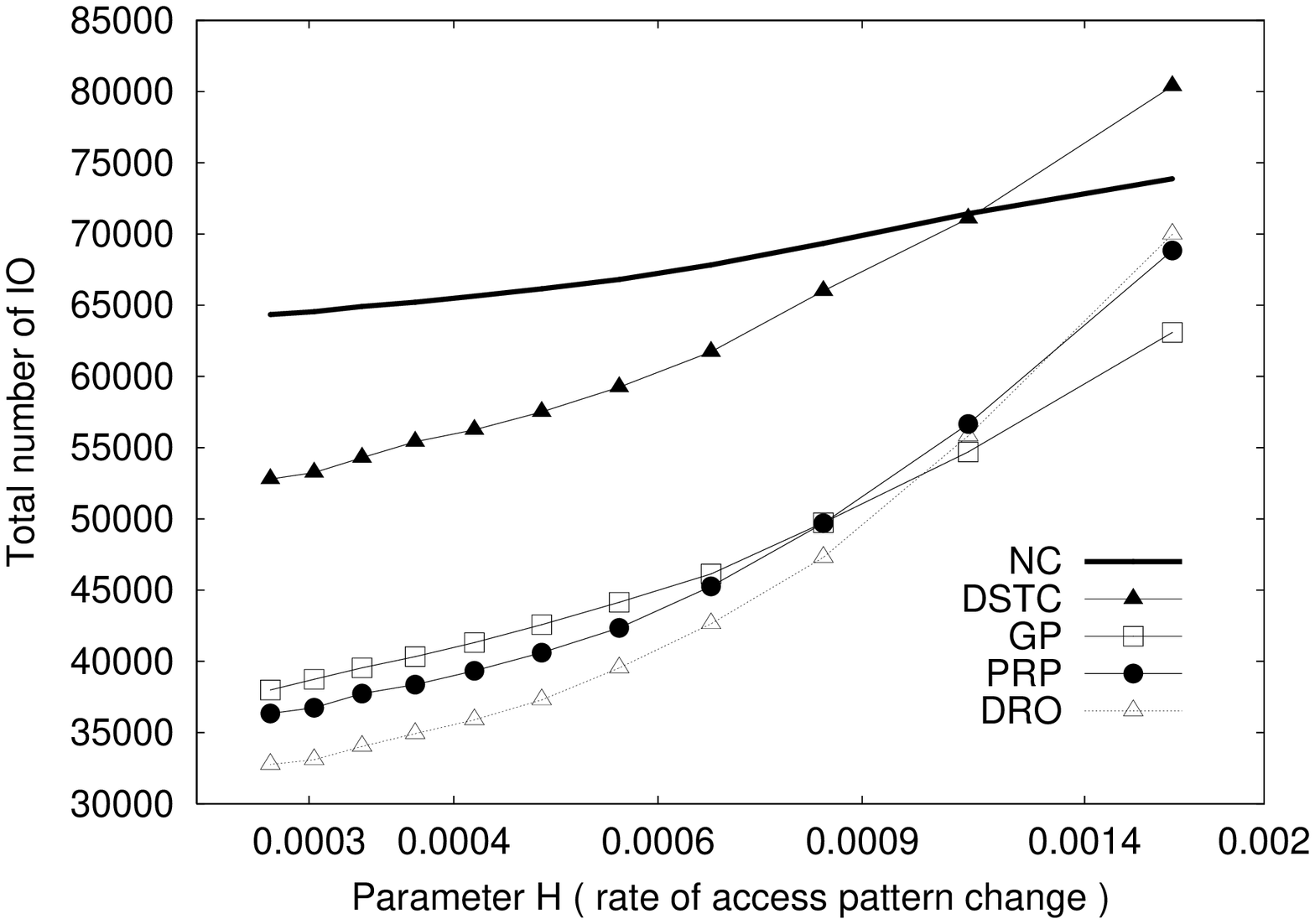}}\quad
  \caption{Regional dependency results}
  \label{fig:rand_move}
  \end{center}  
\end{figure*}

In these experiments, we 
tested the dynamic clustering algorithms' ability to adapt to changes in access pattern
by varying the rate of access pattern change (parameter $H$).
The results of these experiments (Figure~\ref{fig:rand_move}) induce three main conclusions.
First, when rate of access pattern change is 
small ($H$ lower than 0.0006), 
all algorithms show similar performance trends.
Second, when the more vigorous style of change is applied (Figure~\ref{fig:rand_move} (a)), all 
dynamic clustering algorithms' performance quickly degrades to worse than no clustering.  Third, when 
access pattern change is very vigorous ($H$ greater than 0.0006),
DRO, GP, and PRP show a better performance trend, implying these algorithms are more robust to access pattern change.  
This is because these algorithms choose only a relatively few pages (the worst clustered) to re-cluster.
This leads to greater robustness.  We term this flexible conservative re-clustering.  In contrast, DSTC 
re-clusters a page even when there is only small potential gain.  This explains DSTC's poor performance 
when compared to the other algorithms.  
\vspace{-0.5cm}

\vspace{-0.5cm}

\subsection{Moving and Gradual Moving By Reference Experiments}
\label{sec:s-ref_results}
\begin{figure*}[hbt]
  \begin{center}
  \subfigure[Moving Window of Change]{\includegraphics[width = 5.85 cm]{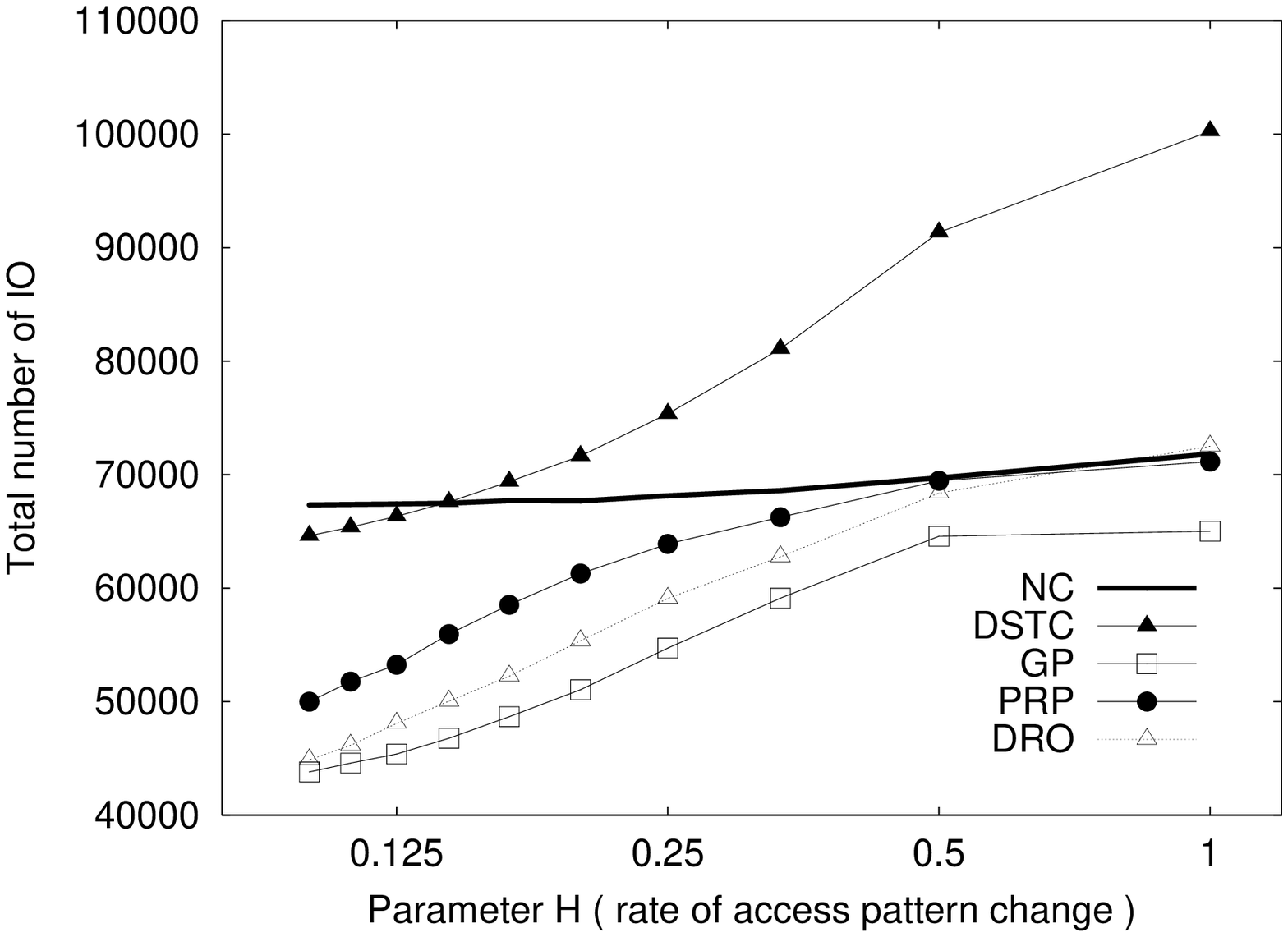}}\quad
  \subfigure[Gradual Moving Window of Change]{\includegraphics[width = 5.85 cm]{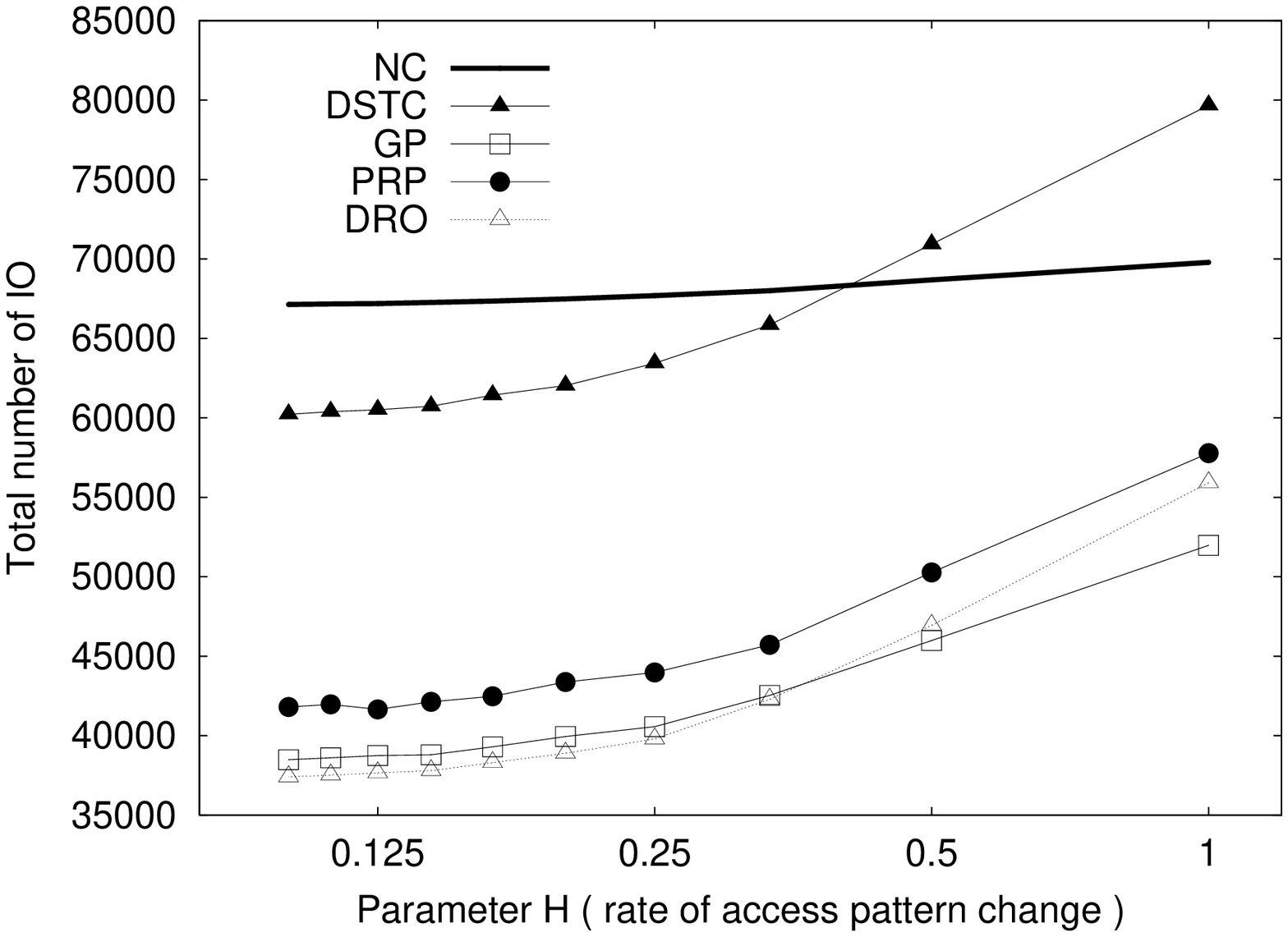}}\quad
  \caption{S-reference dependency results}
  \label{fig:sref_move}
  \end{center}
\end{figure*}

These experiments, 
used the integrated regional dependency protocol method outlined in 
Section~\ref{sec:reg_dep_prot} to integrate \emph{by reference dependency} with the \emph{moving} and 
\emph{gradual moving window of change regional} protocols.  We also used the hybrid 
dependency setting detailed in Section~\ref{sec:hybrid}.  
The random function we used in the first phase
partitioned the database into one hot (3\% database 
size and  80\% probability of reference, which represents typical database application behaviour)
and one cold region.  
The results for these experiments are shown on Figure~\ref{fig:sref_move}.  In the moving window of change results 
(Figure~\ref{fig:sref_move} (a)), DRO, GP and, PRP were again more robust to changes in access pattern 
than DSTC.  However, in contrast to the previous experiment, DRO, GP, and PRP never perform worse than NC by much, 
even when parameter $H$ is 1 (access pattern changes after every transaction).  The reason is the 
cooling and heating of references is a milder form of access pattern change than the pure moving 
window of change regional protocol of the previous experiment.  As in the previous experiment, 
all dynamic clustering algorithm show approximately the same performance trend for the gradual moving 
window of change results (Figure~\ref{fig:sref_move} (b)).

\vspace{-0.5cm}

\vspace{-1.5cm}

\section{Conclusion}
\label{sec:conclusion}

In this paper we presented a new framework for object benchmarking, DOEF,
which allows ODBMSs' designers and users to test  the performances of a given system in a dynamic setting.   This is an important contribution since almost all real world applications exhibit access pattern changes, but no existing benchmark attempt to model this behavior.  

We have designed DOEF to be extensible along two axes. First, new styles of access pattern change can be defined, through the definition of H-regions.  We encourage other researchers or users to extend DOEF by making the DOEF code freely 
available for download\footnote{\it http://eric.univ-lyon2.fr/$\sim$jdarmont/download/docb-voodb.tar.gz}. Second,
we can apply the concepts developed in this paper to object-relational databases. This is made easier by the fact OCB (the layer below DOEF) can be easily adapted to the object-relational context\footnote{Even if extensions would be required, such as abstract data types or nested tables.}.




Experimental results have demonstrated DOEF's ability to meet our objective of
exploring the performance of databases within the context of changing patterns of data access.  
Two new insight were gained: dynamic clustering algorithms can cope with moderate levels of access pattern change but performance rapidly degrades to be worse than no clustering when vigorous styles of access pattern change is applied; and flexible conservative re-clustering is the key in determining a clustering algorithm's ability to adapt to changes in access pattern. 

This study opens several research perspectives.  
The first one concerns the exploitation of DOEF  to keep on aquiring knowledge about the dynamic behavior  of various ODBMSs.  Second, adapting OCB and DOEF to the object-relational model will enable performance comparison of object-relational DBMSs.   Since OCB's schema can be directly implemented within an object-relational system, this would only involve adapting existing and proposing new OCB operations relevant for such a system.  Lastly, the effectiveness of DOEF at evaluating other aspects of database performance could be explored.  Optimisation techniques, such as buffering and prefetching could also be evaluated.



\nopagebreak





\bibliography{paper}

\end{document}